\documentclass[12pt,a4paper]{article}

\usepackage[english]{babel}

\usepackage[letterpaper,top=2cm,bottom=2cm,left=3cm,right=3cm,marginparwidth=1.75cm]{geometry}

\usepackage{amsmath}
\usepackage{graphicx}
\usepackage[colorlinks=true, allcolors=blue]{hyperref}

\usepackage{listings} 
\usepackage{array} 
\usepackage{xcolor}
\usepackage{adjustbox}

\usepackage{tabularx} 
\usepackage{booktabs} 

\definecolor{codegray}{rgb}{0.95,0.95,0.95}
\definecolor{commentgray}{rgb}{0.4,0.4,0.4}
\definecolor{keywordblue}{rgb}{0.2,0.2,0.7}
\definecolor{stringred}{rgb}{0.8,0.1,0.1}
\definecolor{linenumbergray}{rgb}{0.5,0.5,0.5}
\definecolor{framegray}{rgb}{0.75,0.75,0.75}

\lstdefinelanguage{Solidity}{
  morekeywords={
    contract,function,modifier,event,enum,struct,if,else,while,for,import,return,
    mapping,address,bool,string,public,private,internal,external,view,pure,storage,memory,new,
    require,assert,revert,emit,calldata,override,virtual,constructor
  },
  sensitive=true,
  morecomment=[l]{//},
  morecomment=[s]{/*}{*/},
  morestring=[b]",
}

\title{Security Analysis of Ponzi Schemes in Ethereum Smart Contracts}

\author{CHUNYI ZHANG\footnotemark[1], QINGHONG WEI\footnotemark[1], XIAOQI LI\footnotemark[2]}
\date{}

\begin{document}
\maketitle

\renewcommand{\thefootnote}{*}
\footnotetext[1]{Chunyi Zhang and Qinghong Wei contributed equally to this work.}
\footnotetext[2]{Authors' Contact Information: Chunyi Zhang, Hainan University, Haikou, China; Qinghong Wei, Hainan University, Haikou, China; Xiaoqi Li, csxqli@ieee.org, Hainan University, Haikou, China.}

\begin{abstract}
The rapid advancement of blockchain technology has precipitated the widespread adoption of Ethereum and smart contracts across a variety of sectors. However, this has also given rise to numerous fraudulent activities, with many speculators embedding Ponzi schemes within smart contracts, resulting in significant financial losses for investors. Currently, there is a lack of effective methods for identifying and analyzing such new types of fraudulent activities. This paper categorizes these scams into four structural types and explores the intrinsic characteristics of Ponzi scheme contract source code from a program analysis perspective. The Mythril tool is employed to conduct static and dynamic analyses of representative cases, thereby revealing their vulnerabilities and operational mechanisms. Furthermore, this paper employs shell scripts and command patterns to conduct batch detection of open-source smart contract code, thereby unveiling the common characteristics of Ponzi scheme smart contracts.
\end{abstract}

\section{Introduction}
In recent years, blockchain technology has experienced rapid development, accompanied by the meteoric rise of virtual currencies. However, the corresponding regulatory and detection technologies have not advanced significantly. Therefore, the development of a comprehensive analysis and detection system has become an urgent priority \cite{ibba2021evaluating}. The proliferation of fraudulent schemes such as Ponzi scams can be attributed to certain vulnerabilities inherent in human nature, characterized by a proclivity for low-risk, high-return investments. Although pornography, gambling, and drugs have deleterious effects on society, they objectively serve as a litmus test for new detection technologies in these fields. In a similar vein, Ponzi schemes can be regarded as the ``pornography, gambling, and drugs'' of blockchain technology, prompting research in related detection and analysis domains. Such illicit desires and activities frequently emerge in cutting-edge technologies, exploiting ordinary investors' limited knowledge to gain unlawful profits \cite{li2024detecting}. 

The most widely recognized application of blockchain is the Bitcoin cryptocurrency \cite{urquhart2016inefficiency}. Bitcoin's most clandestine yet significant function is its use in money laundering. In 2020, a global ``mining craze'' materialized, engendering a widespread phenomenon. Although a minority of internet users were able to profit from these transactions, the majority were deceived. However, following the enactment of a series of laws and regulations that strictly prohibit virtual currency transactions and categorize them as illicit activities, the mining craze gradually subsided \cite{zheng2018blockchain}. The fraudulent activities within the cryptocurrency trading community subsequently came to light, leading to the realization that virtual currencies ultimately function as a conduit for money laundering \cite{lu2025movescanner}.

\begin{figure}
\centering
\includegraphics[width=0.9\linewidth]{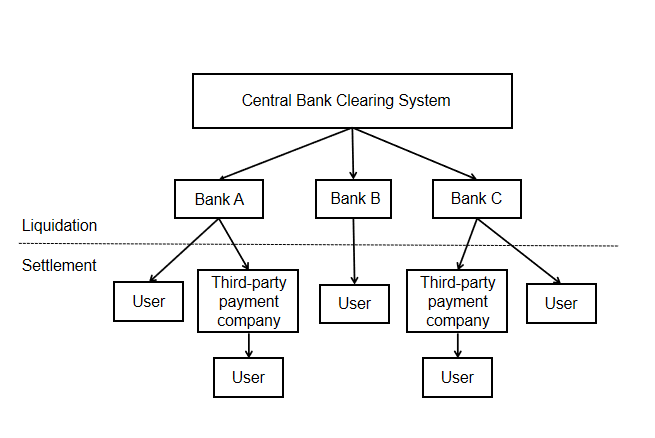}
\caption{\label{fig: Payment System of the People's Bank of China}Payment System of the People's Bank of China}
\end{figure}

Financial institutions have been researching and analyzing blockchain since its emergence \cite{kong2025uechecker}. The fundamental principle of finance is the circulation of capital, while the core function of financial institutions is intermediation, which involves the charging of fees to bridge information gaps between buyers and sellers. The field of finance is predicated on trust. Blockchain's decentralized nature makes it particularly well-suited for the financial sector. Its application can significantly reduce the costs associated with capital flows. Furthermore, the consensus mechanisms that underpin the blockchain are predicated on decentralization, a factor that engenders rapid agreement among any given organization or individual, and thus serves to demonstrate the vast scope of the technology's applicability.  As illustrated in Figure \ref{fig: Payment System of the People's Bank of China}, the payment system of the People's Bank of China comprises several components. The prevailing payment system in China is based on a dual-tier ``settlement-clearing'' model. Transactions settlement is achieved through the collaboration between banking institutions and consumers or merchants, followed by the internal settlement process within the banking system. The processing of interbank clearing is conducted through the central bank's designated clearing system. The application of blockchain technology has the potential to transform this complex workflow into a peer-to-peer transaction model, significantly reducing operational complexity and lowering costs \cite{zhang2025penetration}.

As a valuable complement to traditional internet technologies, blockchain can realize its true potential when fully integrated with technologies such as big data, cloud computing, and artificial intelligence \cite{bharti2019study}. The application of blockchain technology facilitates the redistribution of cloud computing resources, contributing to the reduction of internet operational costs and the safeguarding of user data privacy. Blockchain employs a chained data structure for data validation and storage, utilizes distributed node consensus algorithms to facilitate data generation and updates, ensures secure data transmission and access control through cryptographic techniques, and executes data programming and operations via smart contracts composed of automated script code. This novel distributed infrastructure and computational paradigm provides users with open, traceable, tamper-proof, secure, and reliable distributed ledger services. On the one hand, blockchain's decentralized nature significantly reduces information processing costs and enhances information management efficiency for smart contracts. On the other hand, its consensus mechanisms diminish interpersonal trust costs, enabling machine-based trust. The use of blockchain-based decentralization, anonymity, immutable data, and peer-to-peer transactions facilitates the on-chain recording of data acquisition, exchange, and analysis processes, thus propelling the development of the big data industry. Despite incorporating advanced technologies such as game theory, cryptography, distributed systems, and peer-to-peer (P2P) network protocols to provide numerous conveniences, blockchain technology can also serve as a tool for fraud and money laundering in the hands of malicious actors \cite{peng2025multicfv}.

Ethereum's widespread adoption has enabled Ponzi schemes to hide within smart contracts \cite{zou2025malicious}. The combination of the rapid dissemination of information on the internet and the ease of communication it engenders has rapidly attracted investors seeking to secure rapid financial gains. Blockchain technology is characterized by the maintenance of a growing list of data records through shared nodes. Ponzi schemes hidden within smart contracts promise low risk and high returns, attracting investors with substantial profits. Even those harboring doubts often succumb to the allure of massive gains, and many participants lack even a basic understanding of blockchain concepts or principles. Investors are unlikely to study the source code of smart contracts for investment purposes. Moreover, platforms like Ethereum rarely open-source their code, typically providing only bytecode and transaction data. This makes detecting Ponzi schemes embedded within smart contracts inherently difficult. Furthermore, no legislation or regulations are implemented to penalize such fraudulent activities. The inherent anonymity of blockchain technology provides a security lock for fraudsters, meaning that victims are unable to identify the perpetrators after executing smart contracts. While it is an irrefutable fact that blockchain technology is exceptional, it is equally important to analyze and detect fraudulent schemes such as Ponzi scams if the technology is to develop healthily \cite{zou2025malicious}.

Research on the security analysis of Ponzi scheme smart contracts can be broadly categorized into principles, analysis, detection, and regulation, with most studies focusing on detection \cite{zhang2025risk}. As blockchain technology evolves and finds broader applications, an increasing number of researchers are turning their attention to smart contracts, with security performance emerging as a key area of interest. In response to these security concerns, a significant number of researchers worldwide have conducted in-depth analyses and developed detection schemes for Ponzi scheme smart contracts, aiming to safeguard the information privacy and asset security of Ethereum users. Reports indicate that various Ponzi schemes are profiting heavily from investors who lack extensive knowledge of blockchain technology yet seek high returns. A study of the period between 2 September 2013 and 9 September 2014 revealed that Bitcoin-related scam cases involved losses totaling \$7 million \cite{hinckeldeyn2018short}.

Currently, research on detecting Ponzi scheme smart contracts through program analysis is flourishing. Researchers have proposed various methods and tools to identify potential vulnerabilities and security issues in smart contracts, including code auditing and vulnerability mining using techniques such as symbolic execution and static analysis. In addition, some researchers are developing machine learning-based approaches to detect Ponzi scheme smart contracts, which can learn patterns from large historical datasets and predict future risks \cite{shen2025blockchain}.

Marie Vasek and Tyler Moor categorized fraudulent activities in Bitcoin into four types: mining scams, wallet scams, Ponzi schemes, and fraudulent transactions \cite{vasek2015there}. Massimo Bartoletti et al. were among the earliest researchers to study Ponzi schemes in Ethereum smart contracts \cite{bartoletti2020dissecting}. Based on contract source code collected from etherchain.org, they categorized Ponzi schemes within smart contracts into four types: tree-based Ponzi schemes, chain-based Ponzi schemes, handover-based Ponzi schemes, and cascade-based Ponzi schemes. Addressing smart contract security issues, numerous industry experts and scholars have proposed tools or research methods for vulnerability detection and scam identification. Jiang et al. developed and utilized the ContractFuzzer fuzz testing tool \cite{jiang2018contractfuzzer}. Frank et al. proposed and implemented ETHBMC, a symbolic execution-based boundary fuzz detector, which provides an accurate model of the Ethereum network \cite{frank2020ethbmc}. Gao et al. introduced SMARTEMBED, a deep learning-based similarity detection and code embedding clone detection tool \cite{gao2019smartembed}. It compares code embedding vectors of open-source Solidity smart contracts on Ethereum with known errors, then assists developers in identifying duplicate code and clone-related errors based on similarity between them. Wang et al. developed ContractWard, a machine learning-based vulnerability detection tool for smart contracts \cite{wang2020contractward}. Its workflow involves simplifying smart contract opcodes, extracting binary syntax features, and constructing detection models through sampling and machine learning algorithms. K. Toyoda et al. demonstrated via machine learning that at least 80\% of high-yield accounts can be identified and detected based on Bitcoin transaction frequency, guiding the screening of malicious accounts of potential fraudulent users \cite{toyoda2018multi}. Luu et al. developed Oyente, a smart contract vulnerability mining tool with an automated, three-step workflow \cite{luu2016making}. First, the current global state of Ethereum and the target contract's bytecode are taken as input. Second, the Z3 solver resolves conditional jumps within the contract to identify potential security issues. Third, symbolic paths indicating problematic locations are output to users simultaneously. Tikhomirov et al. proposed SmartCheck, a static smart contract analysis tool \cite{tikhomirov2018smartcheck}. This tool utilizes XPath pattern matching against an XML-based intermediate representation of Solidity code to detect vulnerability patterns within contracts. Feist et al. introduced Slither, a static smart contract analysis tool \cite{feist2019slither}. Tsankov et al. developed Securify, a static contract analysis tool that constructs dependency graphs from bytecode analysis \cite{tsankov2018securify}. It then extracts semantic information from these graphs and compares it against known characteristics to determine vulnerability presence. So et al. developed an automated smart contract vulnerability detection tool focused on arithmetic computations \cite{so2020verismart}. It first uses a generator to transform contract statements into assertions, then employs a verifier to compare these assertions against the contract's invariants, finally determining arithmetic vulnerabilities based on verification results. Boshmaf et al. investigated Ponzi scheme operations by leveraging data obtained from public sources through connection and analysis \cite{boshmaf2020investigating}. This method was applied to analyze MMM, one of the earliest popular Ponzi schemes, focusing on MMM's monetary flows, lifecycle, daily trading volume, external capital movements, fraudulent operations, and geopolitical information. Wu et al. developed trans2vec, a network embedding algorithm capable of extracting features from smart contract addresses \cite{wu2021first}. They used a Support Vector Machine (SVM) classification model to identify phishing nodes and then investigated Ponzi scheme operations by linking and analyzing data obtained from public sources.

\begin{figure}[t]
\centering
\includegraphics[width=0.9\linewidth]{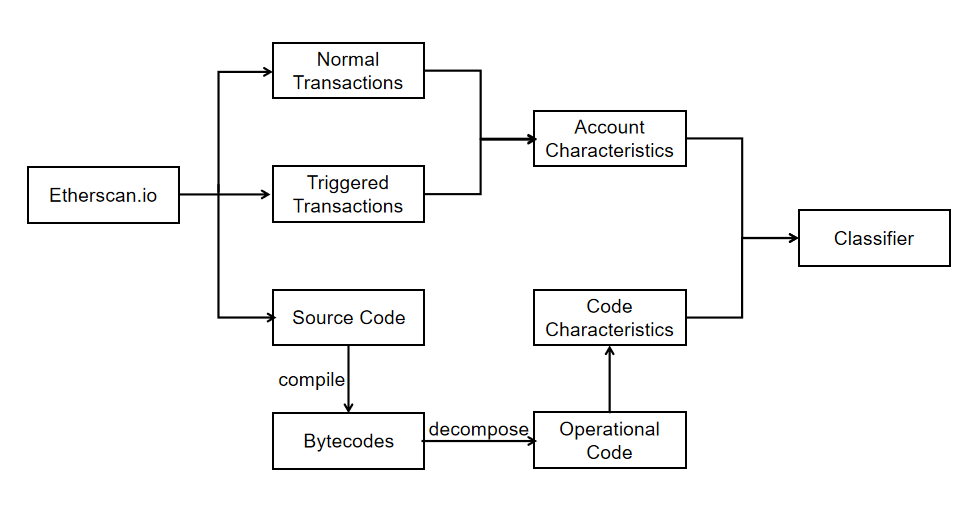}
\caption{\label{fig:Framework for Intelligent Ponzi Scheme Detection Solutions}Framework for Intelligent Ponzi Scheme Detection Solutions}
\end{figure}

Regarding domestic research in this field, ChenWeili et al. proposed using Random Forest (RF) to identify smart contracts \cite{chen2018detecting}. As shown in Figure \ref{fig:Framework for Intelligent Ponzi Scheme Detection Solutions}, account information and transaction source code for smart contracts can be obtained from etherscan.io. The compiled source code yields bytecode, which is then decomposed into opcodes. Finally, a model is constructed based on the frequency of opcode occurrences. Hanyu Xiao employed graph classification methods to detect Ponzi schemes within smart contracts, achieving superior detection performance compared to conventional machine learning approaches. Chunqi Ma proposed a smart contract Ponzi scheme identification method based on global word vector representations. This approach treats smart contract opcodes as words in natural language text, thereby transforming the vectorization problem of code sequences into text vectorization. It successfully identified previously undetected Ponzi scheme contracts \cite{li2021hybrid}.

By analyzing the current state of relevant research both domestically and internationally, we can identify prevailing research hotspots while also discerning existing shortcomings. The existence and concealment of Ponzi schemes within blockchain smart contracts clearly hinder the healthy development of blockchain technology. As blockchain technology advances rapidly, it attracts increasing attention. Consequently, the demand for security analysis of smart contracts on platforms like Ethereum continues to grow, aiming to foster a better and healthier technological platform and trading environment \cite{wang2025ai}.

This study systematically analyzes typical cases of Ponzi scheme smart contracts from a program analysis perspective, examining their operational mechanisms and security vulnerabilities. First, the current state of research on Ponzi schemes in the context of smart contracts is described, with a focus on the foundational theories relevant to this paper. Second, Ponzi schemes are categorized into four structural types based on contract code, and are analyzed both statically and dynamically with specific contract examples. Then, an analysis of 500 open-source contract codes was conducted to identify the common characteristics of Ponzi contracts. Finally, this paper summarizes the research findings and outlines directions for future research improvements.

The main contributions of this study are:
\begin{itemize}
\item \textbf{Analysis of Typical Ponzi Scheme Contracts Based on Structural Classification:}  We conduct a structural classification of Ponzi scheme smart contracts and perform procedural analysis on representative cases within each category, examining their operational mechanisms and critical code.
\item \textbf{Hybrid Static and Dynamic Analysis for Security Testing:} We propose a combined static and dynamic security testing approach, statically auditing the core functionalities and operational mechanisms of contract source code, while dynamically leveraging Mythril to detect code vulnerabilities and trace their root causes.
\item \textbf{Batch Detection of Open-Source Smart Contracts:} We crawl 500 smart contracts and conduct batch inspections using both shell scripts and command patterns, revealing common characteristics among different Ponzi scheme smart contracts.
\end{itemize}

This study aims to provide theoretical support for identifying and detecting smart contracts associated with Ponzi schemes, thereby promoting the healthy development of the blockchain ecosystem.

\section{Background}
When it comes to blockchain technology, it seems like an old topic. However, in recent years, virtual currencies based on blockchain technology have proliferated worldwide, reinvigorating their popularity. However, for the majority of investors, their comprehension of blockchain technology remains largely theoretical, or even confined to its appellation without a comprehensive grasp of its essence. Unscrupulous individuals have been known to exploit this gap by taking advantage of the public's limited understanding of blockchain technology while capitalizing on their eagerness to invest. This exploitation often leads to the perpetration of fraudulent schemes \cite{li2021hybrid}.

\subsection{Blockchain}
In 2008, a scholar using the pseudonym Satoshi Nakamoto published the seminal paper Bitcoin: A Peer-to-Peer Electronic Cash System. The paper proposed creating an electronic payment system enabling transactions between parties without requiring prior knowledge of each other or third-party intervention. This system is grounded in cryptographic principles rather than trust. This seminal paper is widely regarded as the genesis of Bitcoin and a foundational contribution to the evolution of monetary systems on a global scale \cite{peng2025mining}.

From a technical perspective, blockchain is a decentralized, distributed database technology composed of multiple nodes. It employs cryptographic algorithms and consensus mechanisms to ensure data security and credibility. From an application perspective, blockchain can be viewed as a decentralized, distributed database and shared ledger composed of multiple nodes. Its primary characteristics include high security, reliability, and transparency, making it applicable across numerous scenarios such as digital currencies, smart contracts, supply chain management, and the Internet of Things. This ledger facilitates the collaborative maintenance of data by multiple untrusted participants through consensus mechanisms. Once transaction records are generated, they cannot be denied or tampered with by anyone. In recent years, the development of blockchain technology has continued apace, with its applications now encompassing a wide range of sectors, including finance, healthcare, and the Internet of Things \cite{xiang2025security}.

\subsubsection{Ethereum}
Ethereum was founded in late 2013 by Russian-Canadian engineer Vitalik Buterin \cite{choi2021performance}. Ethereum is an open-source, decentralized distributed computing platform built on a public blockchain, where smart contracts can operate \cite{buterin2014next}. It provides the Ethereum Virtual Machine (EVM), a Turing-complete virtual machine. The EVM endows any individual with the capability to develop decentralized applications and smart contracts. Upon initial observation, smart contracts bear a resemblance to conventional programs, as both consist of logical and data blocks.

The native cryptocurrency on the Ethereum platform is Ether (ETH), which can be used for account-to-account transfers or paying for contract resources \cite{li2025interaction}. The pricing mechanism for Ethereum is determined by the interplay between Gas and Gas Price. Gas is defined as the ``work'' required to complete a specific operation, and the cost paid to the network to execute that operation is this ``work'' itself. Therefore, gas can be regarded as the unit consumed per transaction. Gas is a commission paid to miners, settled in ETH, and Gas Price represents the cost per unit of Gas. When a user initiates a transfer on the Ethereum blockchain, the transaction is bundled and uploaded to the blockchain to be finalized. Given the computational demands of this process, a fee is necessary to compensate the miners for their services. This fee is referred to as the Gas Fee. In other words, the Ether paid per transaction constitutes the miner's fee, calculated using Gas and the current Gas Price. Each transaction is subject to a maximum Gas limit, defined as the maximum amount of Gas the sender is willing to expend for that particular transaction. This maximum amount is referred to as the Gas Limit. After the transaction completes, any unused Gas is returned to the sender's account. Users must set the Gas Limit above a certain threshold. If set too low, the transaction is deemed invalid and canceled due to insufficient Gas. Moreover, Gas consumed during computation is not refunded to the sender. It's important to note that, regardless of transaction success, the sender always pays the computational cost to miners \cite{li2024detecting}.

\begin{figure}
\centering
\includegraphics[width=0.9\linewidth]{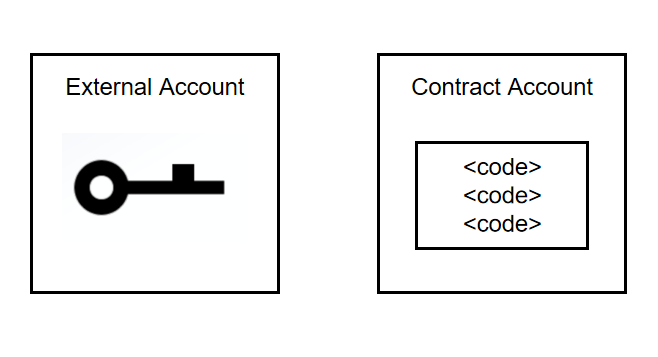}
\caption{\label{fig:Two Types of Ethereum Accounts}Two Types of Ethereum Accounts}
\end{figure}

\begin{figure}
\centering
\includegraphics[width=0.9\linewidth]{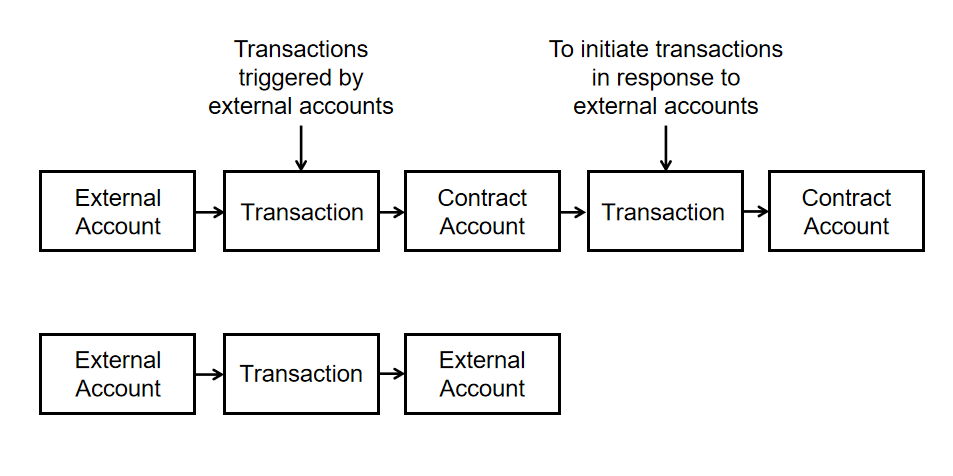}
\caption{\label{fig:The Transaction Relationship Between External and Internal Accounts}The Transaction Relationship Between External and Internal Accounts}
\end{figure}

\begin{figure}
\centering
\includegraphics[width=0.9\linewidth]{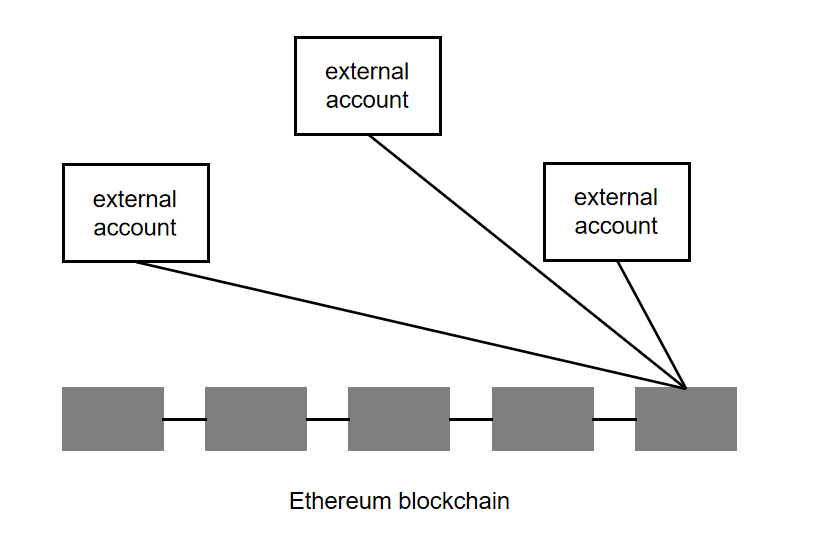}
\caption{\label{fig:The Role of External Account Transactions}The Role of External Account Transactions}
\end{figure}

There are two types of accounts in Ethereum, both sharing the same address space. As shown in Figure \ref{fig:Two Types of Ethereum Accounts}, there are two primary categories of Ethereum accounts: external accounts, which are controlled by a private key space and lack associated code, and smart contract accounts, which are controlled by code embedded within the account. As illustrated in Figure \ref{fig:The Transaction Relationship Between External and Internal Accounts}, contract accounts differ from external accounts in that they cannot actively initiate new transactions but can only trigger transactions in response to others. Therefore, as shown in Figure \ref{fig:The Role of External Account Transactions}, any operation that occurs on the Ethereum blockchain is always processed by transactions initiated by externally controlled accounts. Each Ethereum account consists of an associated state and a 20-byte address \cite{atzei2017survey}. Ethereum addresses identify any account and comprise a unique identifier, data storage, the current Ethereum balance, and a nonce \cite{buterin2014next}. The nonce is a counter that ensures each transaction is executed only once.

\subsubsection{Blockchain Infrastructure}
Structurally, a blockchain can be viewed as a chained data structure composed of multiple blocks. The first block in the blockchain is called the genesis block. Every block in the blockchain, except for the genesis block, has a parent block. Adjacent blocks are linked by referencing the hash value of their parent block \cite{niu2025natlm}.

\begin{figure}
\centering
\includegraphics[width=0.9\linewidth]{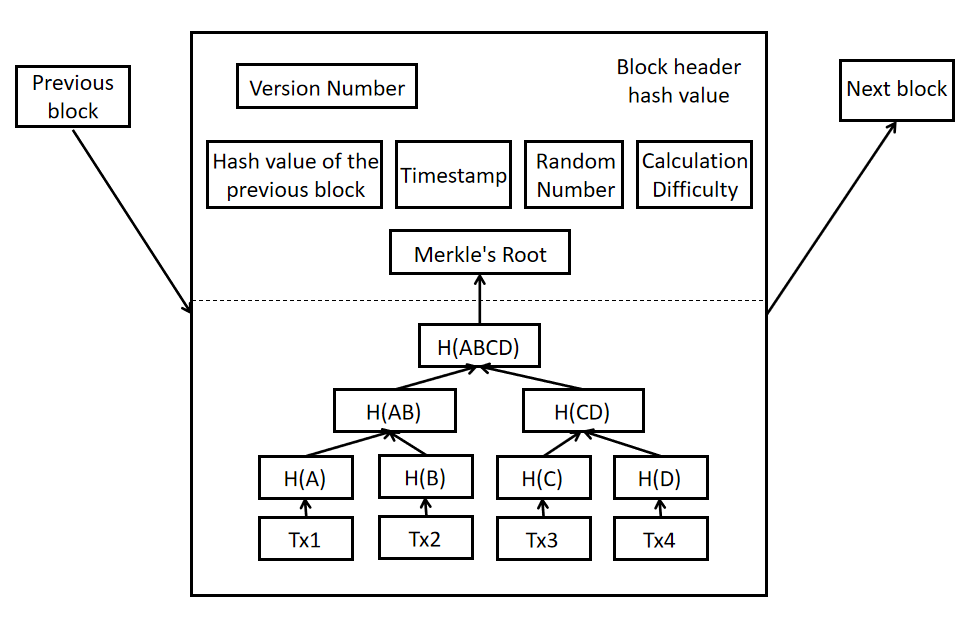}
\caption{\label{fig:Blockchain Diagram}Blockchain Diagram}
\end{figure}

As shown in Figure \ref{fig:Blockchain Diagram}, this is a complete transaction record within a blockchain, illustrating the relationships between its various elements. All transaction information is stored within data blocks Tx. In the diagram, Tx1, Tx2, Tx3, and Tx4 undergo sequential hash computations to derive hash values, which are then stored in the leaf nodes of the Merkle tree. Subsequently, the underlying hash values undergo further pairwise hash computations and are preserved at the root node of the Merkle tree, also known as the root hash value of the Merkle tree.

\begin{figure}
\centering
\includegraphics[width=0.9\linewidth]{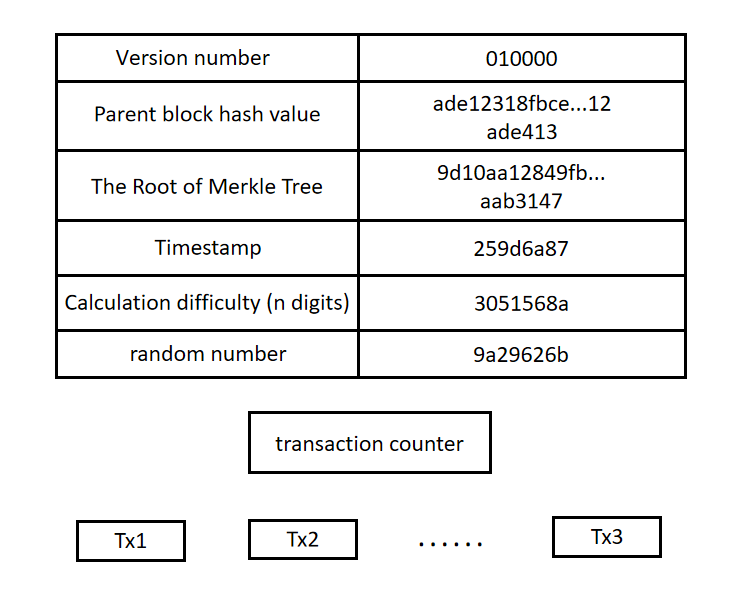}
\caption{\label{fig:Block Structure Diagram}Block Structure Diagram}
\end{figure}

As shown in Figure \ref{fig:Block Structure Diagram}, it shows the contents of various elements within a block following the genesis block. The data in the figure is randomly generated for illustrative purposes and does not represent real data.

As illustrated in Figures \ref{fig:Blockchain Diagram} and \ref{fig:Block Structure Diagram}, a block consists of a block header and a block body. The block header stores:
\begin{itemize}
\item \textbf{Version number:} 4 bytes in size, indicating the block's version and specifying the validation rules it follows.
\item \textbf{Merkle tree root hash:} 32 bytes in size, recording the hash values of all transactions within the block.
\item \textbf{Previous block hash:} 32 bytes in size, recording the parent block's hash address.
\item \textbf{Timestamp:} 4 bytes in size, recording the block's creation timestamp relative to January 1, 1970, 00:00 UTC.
\item \textbf{Computational difficulty:} 4 bytes in size, representing the current difficulty value in compact format.
\item \textbf{Nonce:} 4 bytes in size, recording the computational parameter used to prove work done. It typically starts at zero and increases after each hash calculation is performed.
\end{itemize}

The block body primarily consists of transaction information and transaction counters. The Merkle binary tree structure stores transaction data within the blockchain to ensure that it cannot be tampered with. This is because the root hash value of the Merkle tree is derived by successively hashing transactions. Any alteration to transaction data would consequently change the root hash value \cite{zhou2025blockchain}.

\begin{figure}
\centering
\includegraphics[width=0.9\linewidth]{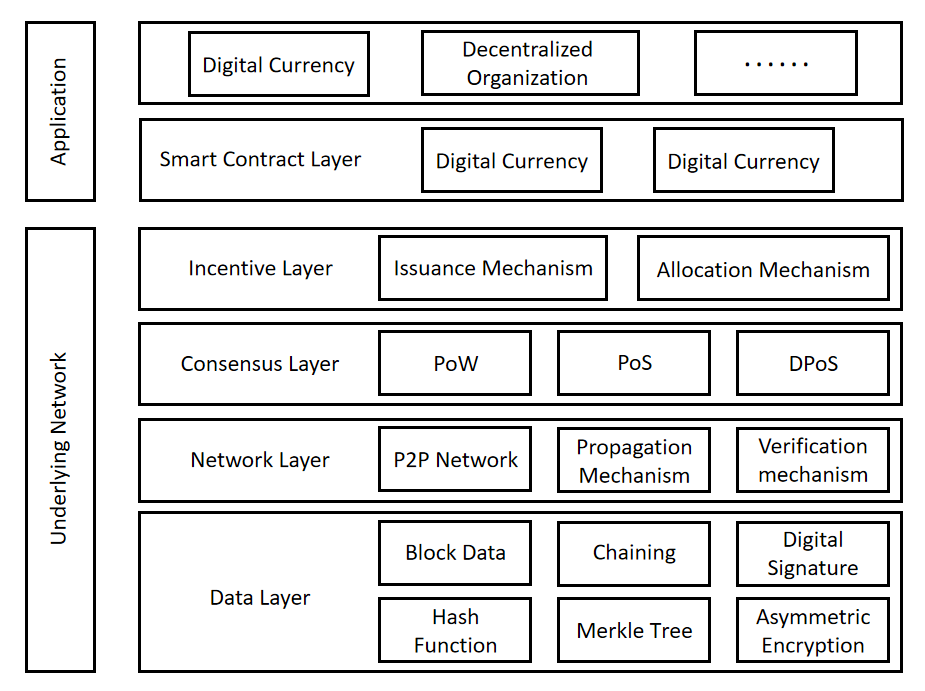}
\caption{\label{fig:Blockchain System Architecture Diagram}Blockchain System Architecture Diagram}
\end{figure}

The system architecture of a blockchain primarily consists of the underlying network and applications. Figure \ref{fig:Blockchain System Architecture Diagram} illustrates the system architecture of a blockchain. As is well known, blockchain is a distributed and decentralized network ledger, so its system architecture aligns with the fundamental characteristics of a network. This network ledger forms a system in which the most fundamental layer is its data structure. This structure organizes information and data in specific formats and methods, inputs them into the blockchain system, and processes them. Once data is input in a standardized manner, the network layer initiates connections, broadcasting and validating information across all nodes. Subsequently, the consensus layer achieves network-wide agreement to create blocks. These blocks can then be used to construct operational platforms for various system products, such as private and consortium chains. Digital currencies, smart contracts, and decentralized organizations are all products that run on blockchain platforms \cite{jin2025blockchain}. 

\subsubsection{Characteristics of Blockchain}
\indent\indent\textbf{(1) Decentralization:} This refers to the fact that the blockchain network structure lacks a centralized institution or individual to control and manage it. Instead, it is composed of numerous nodes that collaborate through protocols to form a distributed network. Within the blockchain network, each node can participate in data verification and transaction processing, meaning that there is no single centralized entity or individual controlling the network's operation and maintenance. This decentralized nature makes blockchain networks more democratic, self-governing, and trustless, thus enhancing network security and credibility.

\textbf{(2) Immutability:} Transaction information for every user in the blockchain is stored and updated across all nodes. Modifications on a single node are invalid, ensuring high data security and stability.

\textbf{(3) Anonymity:} Data interactions between nodes on a blockchain do not require mutual trust, facilitating the accumulation of credibility. Moreover, due to its decentralized nature, blockchain eliminates the need for third-party institutions to store private information.

\textbf{(4) Traceability:} The traceability of blockchain stems from its decentralized and distributed characteristics. Every node on the blockchain can record and verify transactions and data across the network. Consequently, anyone can trace the origin and flow of a specific transaction or piece of data by querying the blockchain.

\subsubsection{Classification of Blockchain Systems}
Blockchains can be categorized into three types based on participants: private blockchains, consortium blockchains, and public blockchains \cite{securities2014us}.

\textbf{(1) Private Blockchain:} Private blockchains have a smaller audience than the other two types. Access permissions are determined by selected nodes within the private chain. Consensus mechanisms are decided by specific organizations within the private chain. Private blockchains can address trust issues and enhance efficiency within responsible organizations.

\textbf{(2) Consortium Blockchain:} A consortium blockchain comprises a group of organizations or entities that have established a cooperative relationship to manage and maintain the blockchain network jointly. Unlike public blockchains, participation in consortium blockchains is limited and requires authorization. Consensus mechanisms are selected through representative nodes, with Proof-of-Stake (PoS) or Practical Byzantine Fault Tolerance (PBFT) commonly employed. Consortium blockchains demand higher security and performance standards than public blockchains. The Hyperledger project, a renowned initiative, exemplifies this model.

\textbf{(3) Public Blockchain:} A public blockchain is an open and decentralized network accessible to anyone, where all transactions and data are transparent and publicly visible. Its defining characteristics include high decentralization, transparency, immutability, and security. On a public blockchain, all participants can validate and record transactions through validation nodes, ensuring exceptional security and trustworthiness. Public blockchains can be applied in various scenarios, including digital currency transactions, smart contracts, and identity verification. Bitcoin and Ethereum are two prominent public blockchain platforms. The consensus mechanism within public blockchains is collectively determined by all nodes, such as Proof of Work (PoW).

\begin{figure}
\centering
\includegraphics[width=0.9\linewidth]{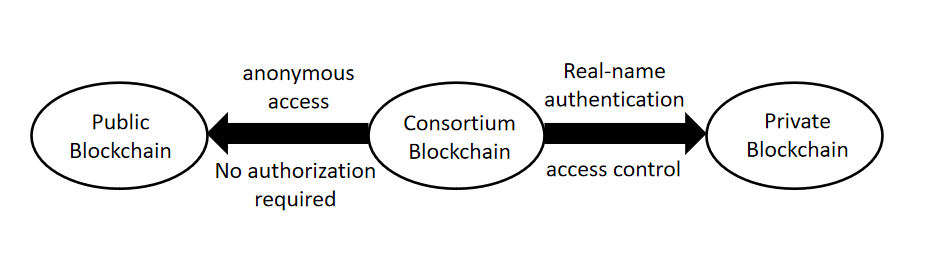}
\caption{\label{fig:Relationships Between Blockchain Systems}Relationships Between Blockchain Systems}
\end{figure}

The relationship among the three is illustrated in Figure \ref{fig:Relationships Between Blockchain Systems}. The comparative relationships among the three are shown in Table \ref{tab:comparison}.

\begin{table}[htbp]
\centering
\caption{Comparison of Different Blockchain Types}
\label{tab:comparison}
\begin{tabularx}{\textwidth}{ 
    l 
    >{\centering\arraybackslash}m{3cm} 
    >{\centering\arraybackslash}m{3cm} 
    >{\centering\arraybackslash}m{3cm} 
}
\toprule[1.5pt] 
\textbf{Property} & \textbf{Private Blockchain} & \textbf{Consortium Blockchain} & \textbf{Public Blockchain} \\
\midrule[0.8pt] 
Read Permission      & Public or Restricted & Public or Restricted & Public \\
\addlinespace[0.3em] 
Centralization       & Yes                  & Partial              & No \\
\addlinespace[0.3em]
Efficiency           & High                 & High                 & Low \\
\addlinespace[0.3em]
Immutability         & Mutable              & Mutable              & Nearly Immutable \\
\addlinespace[0.3em]
Consensus Process    & Permissioned         & Permissioned         & Permissionless \\
\addlinespace[0.3em]
Consensus Mechanism  & Specific Organization & Selected Representative Nodes & All Nodes \\
\bottomrule[1.5pt] 
\end{tabularx}
\end{table}

\subsubsection{Consensus Mechanism}
The consensus mechanism was first proposed in Satoshi Nakamoto's Bitcoin whitepaper. Subsequently, it became one of the core mechanisms in blockchain technology for ensuring data consistency and security. It achieves this through collaboration and validation among multiple nodes, thereby safeguarding the integrity and uniformity of the blockchain. Within a distributed network, data discrepancies may arise across different nodes, necessitating a mechanism to reach consensus and thus guarantee the security and consistency of the entire network.

The origins of consensus mechanisms can be traced back to Byzantine theory \cite{eyal2018majority}, which adheres to two philosophical principles: ``minority obeys majority'' and ``equality among all''. Currently, mainstream algorithms include Proof of Work (PoW), Proof of Stake (PoS), Delegated Proof of Stake (DPoS), and Practical Byzantine Fault Tolerance (PBFT). Each algorithm possesses distinct advantages and disadvantages in practical applications.

\begin{figure}
\centering
\includegraphics[width=0.9\linewidth]{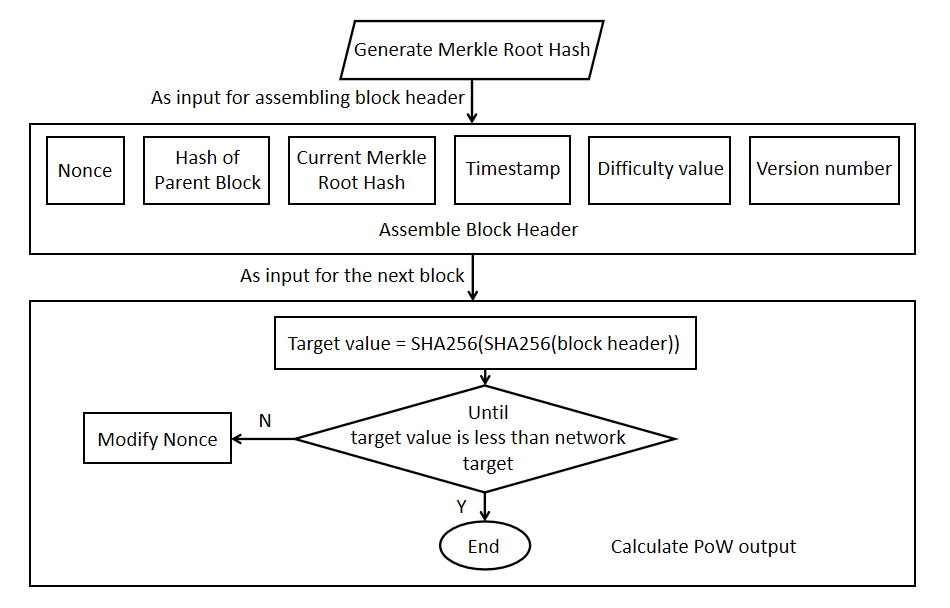}
\caption{\label{fig:Proof of Work Flowchart}Proof of Work Flowchart}
\end{figure}

\textbf{(1) Proof of Work (PoW):} It is commonly used in the Bitcoin network. PoW determines ledger recording rights through computational power competition. Simply put, PoW requires participants to demonstrate sufficient work to prove their trustworthiness. The process where nodes expend computational effort to find qualifying random numbers is termed ``mining''. PoW mining primarily involves three steps:
\begin{itemize}
\item \textbf{Generate the Merkle root hash:} Here, the node generates a transaction to mint tokens on its own.
\item \textbf{Assemble the block header:} The block header serves as an input parameter for calculating the PoW output. It is assembled from the Merkle root hash computed in step one, a random number, and other components.
\item \textbf{Compute the PoW output:} The process is illustrated in Figure \ref{fig:Proof of Work Flowchart}.
\end{itemize}

\textbf{(2) Proof-of-Stake (PoS) Mechanism:} PoS is an improved version of PoW. PoS requires participants to deposit a certain amount of tokens on the blockchain beforehand, somewhat similar to bank savings. It distributes corresponding interest to users based on the quantity and duration of their held digital currency. Placing deposits on the blockchain can incentivize users to make rational decisions. Additionally, reward and penalty mechanisms can be introduced to enhance node operation control, thereby better preventing attacks. Compared to PoW, PoS is more energy-efficient and effective. However, due to extremely low mining costs, it faces a higher risk of attacks. Many blockchains initially adopted PoW but later transitioned to PoS.

\textbf{(3) Delegated Proof of Stake (DPoS):} As a variant of PoS, DPoS inherits some of its advantages and disadvantages. DPoS requires token holders to elect representatives to operate the network, utilizing professionally managed servers to ensure blockchain security and performance. Users can remove underperforming representatives through periodic voting cycles. This approach significantly reduces energy consumption.

\begin{figure}
\centering
\includegraphics[width=0.9\linewidth]{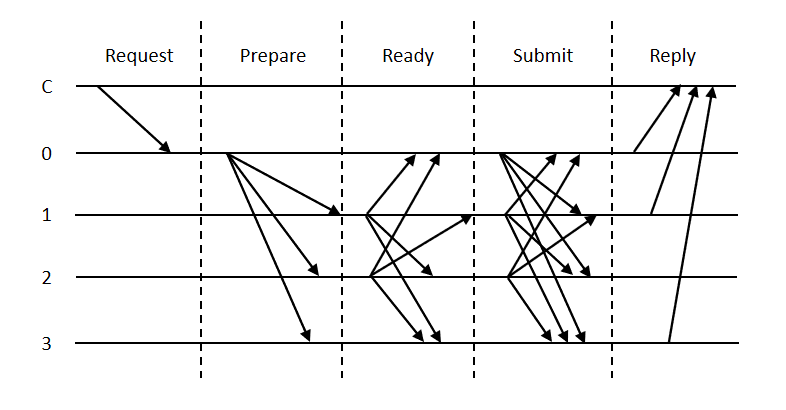}
\caption{\label{fig:Schematic Diagram of the PBFT Consensus Mechanism}Schematic Diagram of the PBFT Consensus Mechanism}
\end{figure}

\textbf{(4) Practical Byzantine Fault Tolerance (PBFT) Consensus Mechanism:} As shown in Figure \ref{fig:Schematic Diagram of the PBFT Consensus Mechanism}, the PBFT message processing path consists of request, pre-prepare, prepare, commit, and reply. The PBFT algorithm relies on cryptographic techniques to ensure tamper-proof message transmission between nodes.

Let $F$ denote the number of invalid or malicious nodes tolerated by PBFT. To maintain system operation, $2F + 1$ valid nodes are required. The total number of nodes in the system is $|R| = 3F + 1$. Therefore, PBFT can tolerate nearly one-third malicious Byzantine replicas, which is precisely why Hyperledger selected PBFT as its consensus mechanism. This also means that at each message processing phase, a node advances to the next phase upon receiving votes from more than two-thirds of the nodes. However, this mechanism requires every network participant to know every node, significantly reducing system throughput and operational efficiency when the number of nodes becomes excessive. PBFT lacks a reward mechanism, and node maintenance consumes Gas, leading to low node participation rates.

\subsection{Smart Contract}
The concept of smart contracts was first proposed by Nick Szabo in 1995, though it remained merely theoretical at the time. Currently, smart contracts have not yet been legally recognized as formal contracts. A smart contract is defined as an automated agreement that is based on blockchain technology. Smart contracts are pieces of computer code that run on the blockchain and manage and enforce the terms of the contract. Smart contracts can automatically execute and enforce the terms and conditions within a contract without requiring a central authority or human intervention, thereby achieving decentralized, automated, and programmable contract execution~\cite{wang2023temporal}. Smart contracts can perform various functions, such as issuing and trading digital currencies, asset management, supply chain management, bill settlement, voting, crowdfunding, and more. This provides more flexible and efficient solutions for diverse financial and non-financial applications.

Within Ethereum, smart contract code is stored on the blockchain as machine code. When a user sends a transaction to the contract, the Ethereum Virtual Machine (EVM) executes the contract's code. Each contract execution consumes computational resources, measured by Gas, with varying Gas requirements per instruction. In addition, users must pay computational resource fees to execute the contract, covering the computational resources consumed during execution. Smart contracts are traceable and non-repudiable.

\subsection{Ponzi Scheme}
The United States Securities and Exchange Commission (SEC) offers the following definition of a Ponzi scheme: A Ponzi scheme is a fraudulent investment scheme involving the payment of purported returns to existing investors from funds provided by new investors.

A Ponzi scheme refers to investment fraud in the financial sector, considered the precursor to pyramid schemes. Many illegal multi-level marketing organizations exploit this method to amass wealth~\cite{wu2024semantic}. This fraudulent practice was originally devised by a speculator named Charles Ponzi. Simply put, it involves using funds from new investors to pay returns to existing investors, creating a false illusion of quick, substantial profits to lure further investments.

Organizers of Ponzi schemes typically attract investors by exaggerating a project's potential, promising high returns with minimal risk. However, since no actual project is implemented, the scheme inevitably collapses. Today, emerging technologies applicable to finance, such as smart contracts on Ethereum, provide fraudsters with new avenues to execute Ponzi schemes. These tools significantly enhance concealment, making it extremely difficult for investors to guard against such scams.

Ponzi schemes disguised as smart contracts typically exhibit the following characteristics:
\begin{itemize}
\item \textbf{Automated execution:} Smart contracts can automatically execute code, making Ponzi scheme operations harder to detect and prevent.
\item \textbf{Anonymity:} Smart contracts can facilitate transactions using anonymous digital currencies such as Bitcoin, making fraudsters harder to track and punish.
\item \textbf{High programmability:} Smart contracts are highly programmable tools that allow fraudsters to easily write, test, and deploy scam code.
\item \textbf{Social media amplification:} Smart contracts can be readily promoted and advertised through social media platforms, drawing broader participation.
\item \textbf{Irreversibility:} Once executed, smart contracts are public, irreversible, and permanently operational. Given these circumstances, investors may mistakenly believe that these false claims are legitimate. Consequently, once defrauded, victims cannot recover losses by canceling transactions or voiding contracts.
\end{itemize}

The causes behind these characteristics are multifaceted and complex. The rapid growth of platforms like Ethereum smart contracts and the permanent nature of smart contracts constantly attract investors seeking substantial profits. Due to the high learning curve of these technologies, most investors lack understanding. Furthermore, from a legal perspective, these technologies remain in a gray area. Therefore, analyzing Ponzi schemes on smart contracts is crucial for developing relevant laws and regulations.

\section{Classification and Program Analysis}
\subsection{Mythril} 
Mythril is a free and open security analysis tool provided by the Ethereum open-source community. This tool detects security vulnerabilities in smart contracts written in Solidity and enables further in-depth analysis. In addition, it serves as a blockchain application security analysis engine for Ethereum smart contracts. Mythril detects various security vulnerabilities in smart contracts through symbolic execution and SMT-based analysis. The main vulnerabilities that can be detected include timestamp dependencies, integer removal, and re-entry attacks.

This experiment involves setting up the Mythril tool environment using an Ubuntu virtual machine. The steps are as follows:
\begin{itemize}
\item \textbf{Install Docker:} Install Docker on a Ubuntu Virtual Machine.
\item \textbf{Check the Docker version:} Check the Docker version to ensure successful installation.
\item \textbf{Start the Docker service:} Start the Docker service to prepare for Mythril installation, and verify if it has started successfully.
\item \textbf{Install Mythril:} Finally, install Mythril using its official Docker image.
\end{itemize}

\subsection{Ponzi Schemes}
Initially, Ponzi schemes lacked a specific classification. However, when combined with smart contract code, the logic of the scam manifests through distinct structural characteristics, giving rise to different types of Ponzi scheme smart contracts. This is precisely the work undertaken by Massimo Bartoletti et al., who categorized Ponzi schemes into four types based on the source code used to deploy their smart contracts.

\subsubsection{Tree-Based Ponzi Scheme}
Tree-based refers to smart contracts whose source code employs tree data structures to store investor addresses. This tree structure represents the sequential order of investors. Etheramid is a classic example of this type of scam, and the following is its core code.

\begin{lstlisting}[language=Solidity]
function enter(address inviter) public { 
    uint amount = msg.value;
    if ((amount < contribution) || (Tree[msg.sender].inviter != 0x0) || (Tree[inviter].inviter == 0x0)) {
        msg.sender.send(msg.value);
        return;
    }

    addParticipant(msg.sender, inviter);
    address next = inviter;
    uint rest = amount;
    uint level = 1;
    while ((next != top) && (level < 7)) {
        uint toSend = rest/2;
        next.send(toSend);
        Tree[next].totalPayout += toSend;
        rest -= toSend;
        next = Tree[next].inviter;
        level++;
    }
    next.send(rest);
    Tree[next].totalPayout += rest;
}
\end{lstlisting}

In this scheme, each investor has an inviter who serves as their parent node. Investors without an inviter become root nodes, effectively the scheme's originators. To join the scheme as an investor, a user must invest funds and designate an introducer as their parent node. During this investment process, the user is rejected if the investment amount is too low, the introducer does not exist, or the user is already present in the scheme. Upon successful investment, the user becomes a child node of their referrer, with their information inserted into the scheme's tree structure. The investment amount is distributed among its parent nodes and ancestor nodes, and closer ancestors receive larger shares. For each additional layer of separation, the distributed amount is halved. This scheme does not impose restrictions on the number of child nodes. Consequently, the more users successfully participate in the scheme, the higher the profits generated. The profit earned by any individual investor within the scheme depends on the number of people they have invited, i.e., the number of child nodes, and is also closely tied to the investment amount of each node.

\begin{figure}
\centering
\includegraphics[width=0.9\linewidth]{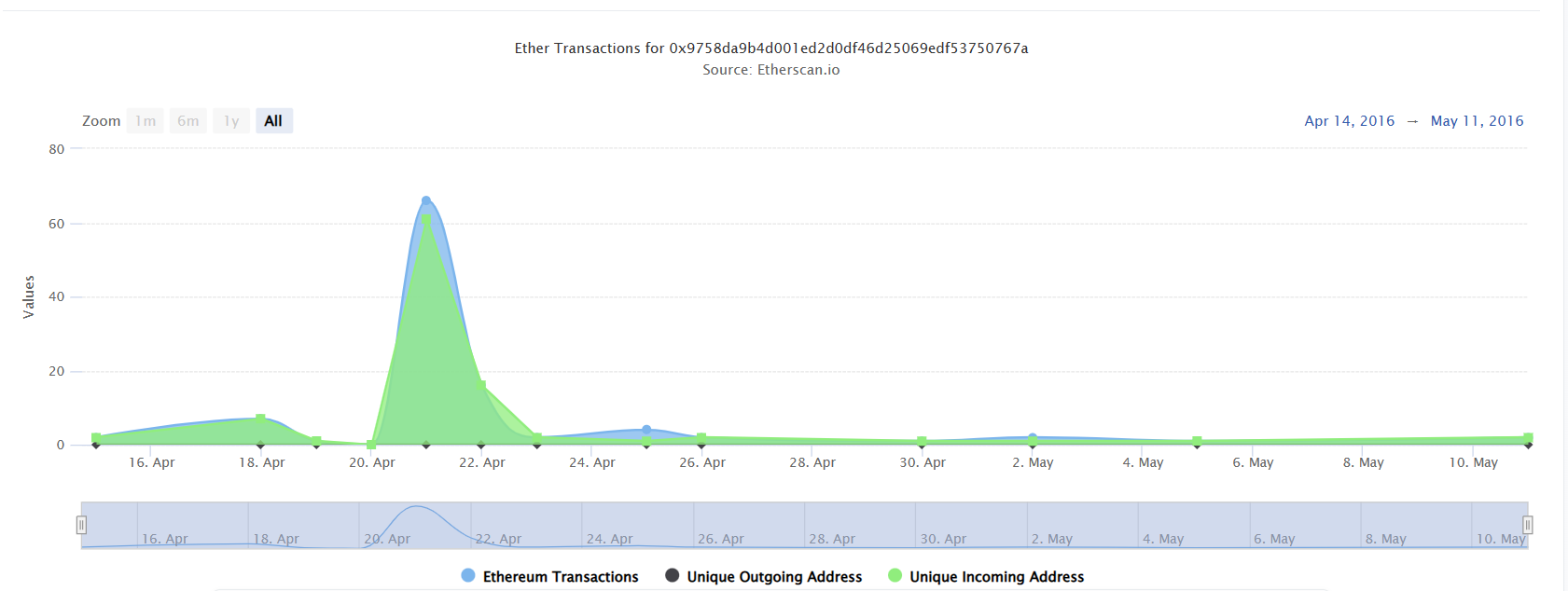}
\caption{\label{fig:Etheramid Contract Trading Activity Chart}Etheramid Contract Trading Activity Chart}
\end{figure}

Based on the above analysis, it is clear that investors generate profits by recruiting new members. Therefore, to maximize profits, it is essential to extend the contract's lifespan as much as possible to attract more investors. As shown in Figure \ref{fig:Etheramid Contract Trading Activity Chart}, the Etheramid contract operated from April 16 to May 11, 2016, with a lifespan of approximately 27 days. Trading activity peaked around April 21. Analysis of its transaction records reveals that 98 accounts were involved in this contract, with 20 accounts achieving a return on investment greater than 1~\cite{wen2024ponzilens+}.

There are two vulnerabilities in the Etheramid contract, both of which are SWC-104, indicating the presence of unchecked call return values. Furthermore, the analysis indicates that the first problematic function is \verb|fallback|. Transactions exploiting this vulnerability consume Gas between 2741 and 37542. 
The consequence of this flaw is that the return value of a message call is not checked, and external calls return a Boolean value. If the callee halts with an exception, 'false' is returned, and execution continues in the caller. The caller should check whether an exception happened and react accordingly to avoid unexpected behavior. For example, it is desirable to wrap external calls in \verb|require()| so the transaction is reverted if the call fails. 
Mythril specifically highlights the problematic line at \verb|msg.sender.send(msg.value)| on line 57, where the return value of the Ether transfer is not checked, creating an SWC-104 vulnerability~\cite{wu2023contraponzi}.
The second issue is found in the \verb|enter(address)| function and is of the same type. In both cases, the contract uses \verb|send()| to transfer Ether without checking the returned Boolean value. If the transfer fails, the contract may continue execution without properly handling the failure, potentially resulting in inconsistent contract behavior.

\subsubsection{Chain-Based Ponzi Scheme}
It is based on array structures, inducing a linear storage sequence among users. This scheme also represents a special case of tree-based Ponzi schemes~\cite{qian2024ds}. Consequently, apart from the storage structure, its other characteristics closely resemble tree-based Ponzi schemes. However, such schemes typically predefine a constant multiplier applied to users' investment amounts, with this multiplier being identical for all participants. This scheme first collects all investments until a predetermined amount is reached, then begins disbursing payments to users, passing the chain structure to the next user.

The CrystalDoubler contract is a classic example of this type of scam, and the following is its core code. User investments trigger the \verb|enter()| function via \verb|msg.value|. The contract requires a minimum investment of 500 fenny. If less than 500 fenny, the contract rejects the user. Upon successful entry, the user's address is stored in the array. After a successful investment, the contract updates its own balance~\cite{liang2024ponziguard}. As new investments continue, the contract constantly checks whether the balance is sufficient to pay the returns to users in the array. If sufficient, the contract pays the current investor double their investment amount. This step loops, paying out to investors awaiting returns until the balance no longer meets the payout condition.

\begin{lstlisting}[language=Solidity]
function enter() { 
    if (msg.value > 500 finney) {

    uint Amount = msg.value;

    // add a new participant to the system and calculate total players
    Total_Players = depositors.length + 1;
    depositors.length += 1;
    depositors[depositors.length-1].EtherAddress = msg.sender;
    depositors[depositors.length-1].Amount = Amount;
    Balance += Amount;              // Balance update
    Total_Deposited += Aomount;     // update deposited Amount
    uint payout;
    uint nr = 0;
    
    while (Balance > depositors[nr].Amount * 200 / 100 && nr < Total_Players) {
        payout = depositors[nr].Amount * 200 / 100;             // calculate pay out
        depositors[nr].EtherAddress.send(payout);               // send pay out to participant
        Balance -= depositors[nr].Amount * 200 / 100;           // balance update
        Total_Paid_Out += depositors[nr].Amount * 200 / 100;    //update paid out amount
    }
    
}
\end{lstlisting}

There are two vulnerabilities within the contract: SWC-101 and SWC-110. SWC-101 indicates an integer data overflow. The issue originates in the \verb|Message()| function, specifically at line 21 of the source code \verb|string public Message = "Welcome|\\ \verb|Player! Double your ETH Now!"|. It is a risk of integer overflow when promising investors to double their returns. SWC-110 indicates a state anomaly. The anomaly occurs in the \verb|depositors(uint256)| function, manifested in line 13 of the source code \verb|InvestorArray[] public depositors|. Specifically, it may trigger an assertion violation because \verb|Solidity's assert()| statement should only be used to check invariants. Review the transaction trace generated for this issue, and either make sure the program logic is correct. Alternatively, if the developer's goal is to restrict user input or enforce preconditions, use \verb|require()| instead of \verb|assert()|. It is important to verify inputs from both the callers (for instance, via passed arguments) and the callees (for instance, via return values).

\subsubsection{Waterfall-Based Ponzi Scheme}
It employs a user ranking similar to the chain-based Ponzi schemes mentioned above, differing only in the logic of currency profit distribution~\cite{qian2024ds}. A chain structure stores each new investor, forming an investor chain. This ensures every investor receives their share when the contract pays out. However, given that payments are made on a first-come, first-served basis, with each distribution commencing from the head of the chain, there is a possibility that investors positioned towards the tail of the chain may never receive profits.

The PonzICO contract is a classic example of this type of scam, and the following is its core code. When total investments fall below 200,000 ETH, the contract first allocates 50\% of incoming funds to the contract owner~\cite{liang2025towards}. The remaining funds are then distributed to prior investors. During this process, if sufficient funds exist to pay the first investor in the chain, that investor receives a proportionate return on their original investment. Finally, the contract continues disbursing funds along the investor chain to the next eligible investor, repeating this process until the remaining balance is insufficient for further payments. There are no obvious vulnerabilities in the contract through Mythril analysis, though there remains the possibility of some deeply hidden or complex vulnerabilities within PonzICO.

\begin{lstlisting}[language=Solidity]
function invest() payable
accreditedInvestor() {
    // first send the owner's modest 50% fee but only if the total invested is less than 200000 ETH
    uint dividend = msg.value;
    uint fee = ownerFee(dividend);
    dividend -= fee;
    
    // then accrue balances from the generous remainder to everyone else previously invested
    for (uint i = 0; i < investors.length; i++) {
        balances[investors[i]] += dividend * invested[investors[i]] / total;
    }

    // finally, add this enterprising new investor to the public balances
    if (invested[msg.sender] == 0) {
        investors.push(msg.sender);
        invested[msg.sender] = msg.value;
    } else {
        invested[msg.sender] += msg.value;
    }
    total += msg.value;
    LogInvestment(msg.sender, msg.value);
}
\end{lstlisting}

\subsubsection{Transfer-Based Ponzi Scheme}
This scheme determines investor entry fees via smart contracts, with fees increasing as new participants join. At any given moment, only one investor receives payouts. Upon receiving payouts, the investor must relinquish their permission to the next user, a process termed permission transfer. To join the scheme, a user must send a certain amount of ETH to the contract to trigger the payout function. The previous user receives this payment from the contract, records the new user's address, and doubles the entry fee~\cite{pennella2025x}.

The PonziScheme contract is a classic example of this type of scam, and the following is its core code. To join the contract, a user must deposit the \verb|startingAmount|, which is the entry fee. The contract transfers this new user's funds to the previous investor, records the new investor's address, and doubles the next entry fee.

\begin{lstlisting}[language=Solidity]
function() payable {
    if(round == 1) {
        if(msg.value != startingAmount) {
            throw;
        }
    } else {
        checkAmount(msg.value);
        
        lastDepositor.send(msg.value);
    }

    lastDepositorAmount = msg.value;
    lastDepositor = msg.sender;
    nextAmount = msg.value * 2;

    increaseRound();
}
\end{lstlisting}

There are three vulnerabilities within the contract: SWC-110, SWC-104, and SWC-110. SWC-110 indicates a state anomaly. It occurs within the \verb|constructor| function, specifically during the call to the function \verb|PonziScheme(uint _startingAmount)| on line 12 of the source code. The specific issue is a triggered assertion violation. SWC-104 indicates that the contract uses an unchecked return value from an external call. This occurs within the \verb|lastDepositor.send(msg.value)| function, meaning the return value generated by this call is not verified. The intended process is: an external call returns a boolean value. If the callee aborts due to an exception, it returns ``false'', allowing execution to resume at the caller. The caller should check for exceptions and respond appropriately to prevent unexpected behavior. For example, wrapping external calls in \verb|require()| is generally advisable, as it allows transactions to roll back if the call fails. Another SWC-110, identical to the one above, but with a different function causing issues. This vulnerability occurs in the \verb|uint public round| definition within the \verb|round| function on line 6 of the source code, also due to triggering an assertion violation.

\subsection{The Fomo3D Ponzi Game}
Fomo3D is a Ponzi scheme game on Ethereum, which is considered to be a form of gambling. The game uses KEY, the token in the Fomo3D game, and the last player to purchase KEY can win a massive prize~\cite{lu2024sourcep}. From a gameplay perspective, and taking into consideration the rewards structure, Fomo3D is transparent in its approach, openly informing players that it functions as a Ponzi scheme, a Ponzi fund. Interestingly, Fomo3D's domain name is exitscam.me, which breaks down to ``exit scam me'', translating to ``leave the scam me''~\cite{feng2024idponzi}. Its smart contract code is open-source, making it a refreshing exception among blockchain gambling games. This approach serves to provide players with the assurance that there are no rigged operators present, nor any clandestine entities that will accept donations and subsequently misappropriate them.

\begin{figure}
\centering
\includegraphics[width=0.9\linewidth]{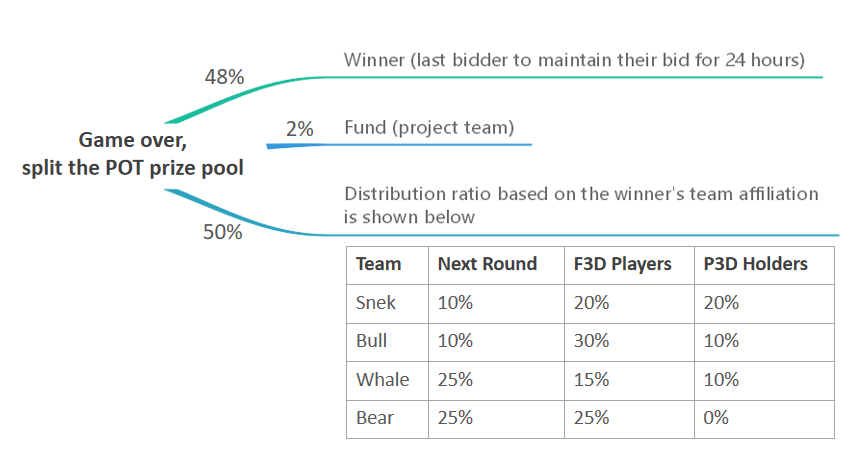}
\caption{\label{fig:Fomo3D Dividend Distribution Ratios Among Different Teams}Fomo3D Dividend Distribution Ratios Among Different Teams}
\end{figure}

In order to participate in Fomo3D, it is first necessary to transfer a specific quantity of ETH to the designated contract address. Following this, the participant is required to select their preferred team from among the options of Snek, Bull, Whale, or Bear~\cite{zhang2022security}. Subsequently, a specific quantity of KEY will be received at the prevailing exchange rate. It has been determined that 86\% of the ETH deposited by participants will be allocated to the prize pool and dividends. However, it should be noted that the distribution ratio varies according to team selection, as illustrated in Figure \ref{fig:Fomo3D Dividend Distribution Ratios Among Different Teams}. A proportion of 10\% is allocated as a form of remuneration to those who successfully refer a new customer to the organization. In the absence of a referrer, the reward is distributed to all KEY holders. A proportion of 2\% is allocated to the game project team, 1\% is designated for Pot Swap, and 1\% is allocated for random airdrops. It is noteworthy that each purchase exceeding 0.01 ETH increases the probability of a random airdrop by 0.1\%.

When someone purchases a pass, a 24-hour countdown to win automatically begins~\cite{onu2023detection}. If no one else buys a pass within that 24-hour period, the last person to purchase a pass becomes the game's winner and receives 48\% of the prize pool. However, if another purchase occurs within those 24 hours, the countdown is reset and a further 30 seconds is added. This winner mechanism invariably compels numerous players to purchase tickets during the final 30 seconds of the countdown, hoping to be the final purchaser. After the advent of Fomo3D, a plethora of analogous games have emerged, all of which require players to purchase game items with ETH. In these games, the triumphant player is entitled to claim the so-called ``grand prize'', while regular participants have the opportunity to obtain ``airdrop rewards''~\cite{galletta2024explainable}.

This game is governed by smart contracts, which means it can never truly be stopped. When Fomo3D first launched in 2018, many feared it would snowball out of control, potentially crashing Ethereum. This concern stemmed from the risk that excessive ETH investment by players seeking victory could flood the Fomo3D pool, causing a critical shortage of circulating ETH on the Ethereum network~\cite{qu2025detecting}. Fortunately, although Fomo3D once reached nearly 30,000 ETH, flaws in the contract code allowed hackers with advanced attack techniques to increase their chances of receiving ``airdrop rewards'' and even use special methods to claim the ``grand prize''. Due to severe exploitation of these two reward mechanisms by hackers, the game's appeal to users diminished, failing to attract new capital. This ultimately created a vicious cycle, rendering it incapable of significantly impacting Ethereum.

Of all funds entering the game, specifically ETH, 1\% flows into the secondary prize pool. The airdrop probability starts at 0\%. Each order exceeding 0.1 ETH increases the airdrop probability by 0.1\%, capped at 1\%~\cite{zheng2023securing}. Players who purchase orders between 0.1 and 1 ETH have a chance to win one-quarter of the secondary prize pool, with higher purchase amounts increasing their probability. The Fomo3D airdrop reward mechanism contained two vulnerabilities. First, the ``random number'' within the contract was not truly random but predictable. Second, there was a flaw in the logic that determined whether the caller was a contract address.

\textbf{(1) ``Random Number'' Vulnerability:} The airdrop reward mechanism relies on a ``random number'' generated internally by the smart contract, with values ranging from 0 to 99. In Fomo3D's source code, the airdrop function \verb|airdrop()| controls this ``random number'' generation. To produce this value, the smart contract performs calculations based on various block information and the transaction initiator's address, making it clearly predictable.

\begin{lstlisting}[language=Solidity]
/**
 * @dev generates a random number between 0-99 and checks to see if thats 
 * resulted in an airdrop win
 * @return do we have a winner?
 */
function airdrop()
    private
    view
    returns(bool)
{
    uint256 seed = uint256(keccak256(abi.encodePacked(

        (block.timestamp).add
        (block.difficulty).add
        ((uint256(keccak256(abi.encodePacked(block.coinbase)))) / (now)).add
        (block.gaslimit).add
        ((uint256(keccak256(abi.encodePacked(msg.sender)))) / (now)).add
        (block.number)

    )));
    if((seed - ((seed / 1000) * 1000)) < airDropTracker_)
        return(true);
    else
        return(false);
}
\end{lstlisting}

\textbf{(2) ``Contract Address Verification'' Vulnerability:} To prevent automated attacks on the contract, Fomo3D developers specifically created the \verb|isHuman()| function to prohibit players from predicting winning random numbers within the contract. Within the \verb|isHuman()| function, the developers made an error. The \verb|extcodesize| operator retrieves the code size at a target address. However, for successfully deployed contracts, their addresses correspond to specific code, causing \verb|extcodesize| to always return a value greater than zero. Consequently, this method could be exploited to determine whether a target address belonged to a contract~\cite{song2022study}. The developers of the Fomo3D game evidently employed a similar approach to prevent contracts from calling specific functions. However, they likely overlooked a glaring vulnerability in this method. By calling the game participation function in the contract construction method, one can bypass the restrictions set by the developer. This vulnerability arises because, during the process of constructing the contract, the contract address does not yet correspond to any code, meaning \verb|extcodesize| returns 0 at this stage.

\begin{lstlisting}[language=Solidity]
/**
 * @dev prevents contracts from interacting with fomo3d
 */
modifier isHuman() {
    address _addr = msg.sender;
    uint256 _codeLength;

    assembly {_codeLength := extcodesize(_addr)}
    require(_codeLength == 0, "sorry humans only");
    _;
}
\end{lstlisting}

By exploiting these two vulnerabilities in combination, attackers can create malicious contracts to participate in the game. Leveraging their ability to predict random numbers, they can significantly boost their win rate and profit with impunity~\cite{chen2024ponzi}.  
The Mythril tool did not detect any obvious vulnerabilities in the Fomo3D contract. This demonstrates that Mythril is not infallible and cannot detect all vulnerabilities. Therefore, when using program analysis methods to detect Ponzi scheme smart contracts, it is advisable to supplement tool-based analysis with source code review.

\section{Batch Detection and Result Analysis}
\subsection{Experimental Data Collection}
The smart contract source code data utilized for batch detection was obtained through a process of data collection from the Etherscan.io website using web crawlers~\cite{cai2023ponzi}. Since the last smart contract flagged as a Ponzi scheme on Etherscan.io predates 2019, all crawled smart contract source code originates from before 2019, making this comparison more meaningful. The scraped data contains 500 smart contracts.

\begin{figure}
\centering
\includegraphics[width=0.9\linewidth]{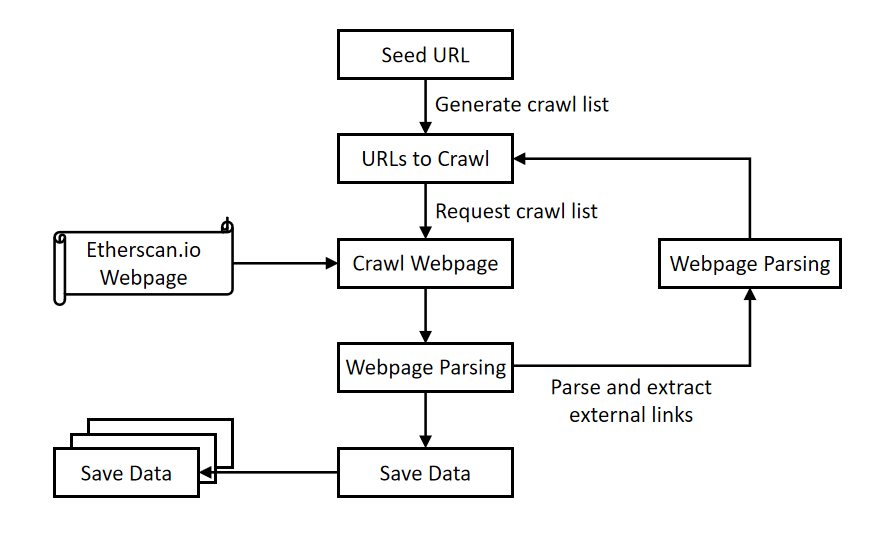}
\caption{\label{fig:Web Crawler Retrieves Sample Data}Web Crawler Retrieves Sample Data}
\end{figure}

The sample data in this paper were obtained through web scraping technology, with the scraping structure illustrated in Figure \ref{fig:Web Crawler Retrieves Sample Data}. The seed URL refers to the starting web address where a crawler program begins its crawl, i.e., the entry point. Typically, a seed URL is a webpage link containing the target data. The crawler program starts from this link and recursively crawls other links contained within it until predetermined termination conditions are met. The URLs to crawl refer to web links discovered by the crawler during the process that have not yet been visited or processed. These are URLs that have been deduplicated from seed URLs and crawled webpages.

When crawling webpages, set the retry mechanism to 25. This primarily prevents data download failures caused by network fluctuations during crawling and improves crawling efficiency. Data parsing refers to the process of extracting target data from formats such as HTML, XML, and JSON~\cite{liu2025ponzi}. After obtaining the webpage content from the URL, simulate browser header information to send messages to Etherscan.io. Finally, save the contract source code retrieved from Etherscan.io as sol files. Name each file according to the crawl sequence, then store all sol files in the code directory.

\subsection{The Batch Detection Process}
Mythril smart contract analysis tools can perform detection and analysis on individual smart contracts as well as multiple smart contracts simultaneously, enabling batch detection. Batch detection in Mythril can be performed in two ways: shell scripts and command mode~\cite{zhang2025security}.

The shell script method operates through a structured workflow. First, it traverses the folder and its subdirectories. Then, it locates files with the ``.sol'' extension within the specified directory. It executes the detection statement to open the directory containing the smart contract requiring inspection, thereby utilizing the Mythril tool to analyze the smart contract. Finally, it saves the terminal detection results to the ``result.txt'' file within the current directory.

\subsection{Results Analysis}
Analysis of bulk smart contract detection results reveals that Ponzi scheme smart contracts operate relatively covertly. The entry fee for investors joining the scheme is typically high, often set with a minimum threshold. Investors below this threshold are denied participation, making a high barrier to entry a defining characteristic of Ponzi schemes within smart contracts.
The scheme's payout structure to investors is determined by the storage mechanism within the smart contract. Typically, the contract initiator reaps the greatest profits, while early investors earn higher returns than later participants.
Smart contracts containing Ponzi schemes are frequently detected by Mythril as harboring integer data overflow vulnerabilities and abnormal state vulnerabilities. The payout structure is one factor contributing to the occurrence of data overflow vulnerabilities in Ponzi scheme smart contracts.

\section{Conclusion and Future Work}
In recent years, with the continuous advancement of digital currency technology, a vast number of new smart contracts are deployed daily on blockchain platforms such as Ethereum, with their quantity constantly reaching new heights. Concurrently, there has been an observed increase in the number of fraudulent transactions that are being concealed. In such circumstances, it is imperative to enhance analytical and detection capabilities in conjunction with vigilance to prevent further individuals from falling victim to scams, thereby fostering a conducive environment for healthy blockchain development.

This paper analyzes and detects Ponzi scheme smart contracts from a programmatic perspective, divided into static code analysis and dynamic analysis using the Mythril tool. Based on the structure of the smart contract source code, Ponzi schemes can be categorized into tree-based, chain-based, waterfall-based, and transfer-based schemes. This paper analyzes these four categories, examining representative smart contracts for each. First, it conducts a static audit of the core functionalities in the contract source code, dissecting their operational principles and comparing them against the characteristics of their respective categories. Second, the Mythril tool is employed to detect vulnerabilities within the contract code and analyze the root causes of these risks. Crawler technology is employed to extract open-source smart contract code from Etherscan.io. Subsequently, shell scripts and command-line interfaces are employed to process the extracted code in batches for analysis via Mythril.

However, the prevailing mainstream approach to analysis and detection currently relies on machine learning to examine Ponzi scheme smart contracts, thereby enhancing detection effectiveness through model training. This paper analyzes the code characteristics of Ponzi schemes within smart contracts from a program analysis perspective. This approach directly correlates theoretical frameworks with practical smart contracts, thereby offering a more profound understanding of the operational mechanisms of various Ponzi schemes within smart contracts. Nevertheless, when compared to machine learning methods, program analysis is more akin to manual labor, entailing significant time and effort.

Consequently, the analysis and detection of Ponzi scheme smart contracts will undoubtedly be enhanced by the adoption of machine learning approaches. This process entails the selection of features to train detection models and their subsequent application to smart contracts. Future research should focus on integrating program analysis with machine learning to more efficiently address the challenge of detecting Ponzi scheme smart contracts.

\bibliographystyle{plain}
\bibliography{bibliography}

\end{document}